\documentclass[a4paper,11pt]{article}
\usepackage{jheppub} 
\usepackage{lineno}
\usepackage{braket}
\usepackage{tensor}
\usepackage{cleveref}
\usepackage{subcaption}

\title{\boldmath Preheated inflation}

 \author[a]{Diogo S. Gorgulho}
 \author[a]{Jo\~ao G. Rosa}
 \affiliation[a]{Univ Coimbra, Faculdade de Ciências e Tecnologia da Universidade de Coimbra and CFisUC, Rua Larga, 3004-516 Coimbra, Portugal}

\emailAdd{diogo.s.gorgulho@proton.me}
\emailAdd{jgrosa@uc.pt}

\abstract{We propose a new mechanism of non-thermal particle production {\it during inflation} based on a narrow parametric resonance, akin to the dynamics of post-inflationary preheating. The mechanism is based on the production of scalar particles with a mass that is an oscillating function of the slowly-rolling inflaton field. This is achieved in a scenario for the collective spontaneous breaking of a U(1) gauge symmetry that, while originally proposed in the context of warm inflation, leads to non-equilibrium particle production sustaining a (sub-dominant) non-thermal radiation bath throughout inflation. We show that this may leave an observational imprint, namely oscillatory features in the primordial curvature power spectrum alongside a (mild) resonant enhancement of its amplitude, as well as secondary gravitational waves that can be probed with future CMB experiments.}

\begin{document}
	\maketitle
	\flushbottom
	
	\section{Introduction}
	\label{sec:intro}
	
		The inflationary paradigm is currently the most compelling solution to the shortcomings of standard cosmology \cite{Guth:1980zm, Linde:2007fr,Linde:1981mu,Linde:1983gd}. In its simplest form, the inflaton scalar field follows a slow-roll trajectory, during which it mimics an approximate cosmological constant, yielding quasi-de Sitter expansion for at least 50-60 $e$-folds.~This framework elegantly explains several features of the presently observable Universe, namely its flatness, isotropy and homogeneity.~Additionally, the quantum fluctuations of the inflaton field generate a nearly scale-invariant spectrum of curvature fluctuations on super-horizon scales, which may give rise to the observed temperature and polarization anisotropies of the Cosmic Microwave Background (CMB) and to the large-scale structure of the Universe. For a pedagogical review of inflation, see Ref. \cite{Riotto:2002yw}.
	
	In the simplest form of the theory, after the slow-roll regime, a stage of particle production known as \textit{reheating} ensues \cite{Kofman:1994rk,Kofman:1997yn,Braden:2010wd}, during which the inflaton (directly or indirectly) decays into Standard Model species while oscillating about the minimum of the scalar potential. It is conventionally assumed that most elementary particles were created during this period, after which standard cosmology is recovered. Reheating is typically divided into three stages. In the first one, dubbed \textit{preheating}, the inflaton field decays into scalar particles via a parametric resonance, normally of the broad (explosive) kind \cite{Kofman:1997yn}. In general, this process is incomplete, as the resonance eventually becomes \textit{narrow} and inefficient due to backreaction effects. Moreover, the resulting particles are far from thermal equilibrium and have large occupation numbers. The second stage is the decay of the previously produced particles and of the remaining inflaton field \cite{Albrecht:1982mp,Dolgov:1982th,Abbott:1982hn}, whose products are then thermalised during the third and final stage of reheating \cite{Kofman:1997yn}.
		
	In the stage of preheating, the classical inflaton field $\phi$, which is oscillating with a decreasing amplitude $\Phi(t)$, couples with~e.g.~a scalar field $\chi$, thus introducing an oscillatory mass term for these particles. Near the minimum the scalar potential is approximately quadratic, $V(\phi )\sim \frac{1}{2}\,m^2(\phi-\sigma)^2$, so that considering an interaction term of the form $-\frac{1}{2}g^2 \phi^2 \chi^2$ and neglecting expansion, the equation of motion for each mode $\chi_k(z)$, with $z = \frac{m\,t}{2}$, can be written as a Mathieu equation \cite{mclachlan1947theory, Kofman:1997yn,NIST}
	\begin{equation}\label{eq:mathieu_eq_preheating}
		\chi_k''(z) \: + \: \left[A_k\:-\:2 q\,\cos(2z)\right]\, \chi_k(z) \:=\:0 \:\:, \quad \begin{cases}
			A_k = 4\, \frac{k^2+g^2 \sigma^2}{m^2} \\
			q = \frac{4g^2 \sigma \Phi}{m^2}
		\end{cases}~.
	\end{equation} 
	\noindent This equation is notable for the fact that its solutions develop parametric resonances (which can either be \textit{narrow}, for $q \ll 1$, or \textit{broad}, for $q \gg 1$) in the form of exponential instabilities $\chi_k(z) \propto e^{\mu_k(z)\,z}$, the instability being attained when $\mu_k(z)$ is real \cite{Rosa:2007dr, Kofman:1997yn, mclachlan1947theory,NIST}. Using these unstable solutions, one can obtain the number of produced particles with momentum $k$ and their total number density:
	\begin{equation}\label{eq:number_density_preheating}
			n_k = \frac{\omega_k}{2}\, \left( \frac{|\dot \chi_k|^2}{\omega_k^2}+ |\chi_k|^2 \right) - \frac{1}{2} ~,\quad\qquad\qquad n_\chi = \int \frac{d^3 k}{(2\pi)^3}\: \:n_k~,
	\end{equation}
	\noindent with $\omega_k^2(t) = k^2 + g^2 \sigma^2 + 2g^2 \sigma \Phi\, \sin(mt)$, so that the exponential growth can be interpreted in this context as explosive particle production (rigorously, the production is explosive only in the case of a broad resonance). Note that preheating is not limited to the decay of the inflaton field into other scalar fields: for example, a theory of preheating with fermions was developed in \cite{Baacke:1998di,Greene:1998nh,Greene:2000ew} and one with Abelian gauge fields was proposed in \cite{Deskins:2013dwa}. The subsequent decay of the particles produced during preheating and of the remaining inflaton field, as well as the thermalisation of their decay products, is described by methods similar to those used in the elementary theory of reheating (see \cite{Kofman:1997yn, Albrecht:1982mp, Abbott:1982hn, Dolgov:1982th}).
	
	In spite of the existence of a post-inflation reheating period, there have been several proposals of particle production mechanisms during the slow-roll phase of inflation \cite{Chung:1999ve,Cook:2011hg,Creminelli:2023aly,Durrer:2023rhc,Barnaby:2009dd, Yu:2023ity}. This is interesting for a variety of reasons. Firstly, the presence of additional fields during inflation should in principle cause some backreaction on the dynamics of the classical inflaton and of its quantum fluctuations, which may leave an observable imprint on CMB anisotropies, thus providing a new window into the high-energy physics model behind inflation. This does not happen in the usual picture of post-inflation reheating, as the perturbations generated in this stage do not grow to macroscopic sizes. Additionally, the produced particles may generate tensor perturbations of their own, leading to a stochastic gravitational wave (GW) spectrum that will contribute to the primordial tensor power spectrum. Moreover, this spectrum might eventually be detectable by current or future GW interferometers \cite{Caprini:2018mtu, Cook:2011hg, Guzzetti:2016mkm,Boyle:2005se}.
	
	Secondly, if particle production during inflation is efficient enough, there may be no need for a separate post-inflationary reheating period, since the inflaton's energy may be fully converted into radiation by the end of the slow-roll regime (assuming, of course, that Standard Model particles are eventually produced by decays and/or thermal scatterings within the produced radiation bath. 
	
	Lastly, particle production may lead to an additional friction that damps the inflaton's motion, therefore facilitating the slow-roll dynamics and relaxing the constraints on the scalar potential's slope and curvature, while also potentially modifying the primordial curvature power spectrum. This is e.g.~the case of warm inflation models \cite{Berera:1995ie,Berera:1995wh,Berera:2008ar, Bastero-Gil:2009sdq}, where a nearly-thermal radiation bath at temperature $T\gtrsim H$ is sustained during inflation by dissipative processes. The thermal nature of inflaton fluctuations in this setup, and their coupling to statistical fluctuations in the radiation bath, also significantly affects the dynamics and spectrum of scalar curvature perturbations.

 Despite its appeal, warm inflation models are challenging to realize within consistent quantum field theory models. Finite-temperature dissipative particle production is only efficient for particle masses $\lesssim T$, while conventional couplings to the inflaton field tend to make the produced particles heavy, $m\sim g\phi$, where $g$ denotes the relevant coupling constant. Thermal backreaction may also reintroduce the ``eta-problem'', since it shifts the second potential slow-roll parameter by an amount $\Delta\eta_V\sim \frac{g^2 T^2}{H^2}$. The additional friction allows for slow-roll with $\eta < 1+\frac{\Upsilon}{3H}$, where $\Upsilon$ is the finite-temperature dissipative coefficient, but this may be insufficient since $\Upsilon \lesssim T$ in all models considered in the literature. In fact, formally the friction coefficient diverges for on-shell particle production with a vanishing thermal width, but this would correspond to an arbitrarily long relaxation time of the radiation bath, which is incompatible with the underlying assumption that a quasi-thermal equilibrium can be maintained during inflation.

A possible avenue to overcome these issues is the Warm Little Inflaton (WLI) scenario and its variants \cite{Bastero-Gil:2018yen, Bastero-Gil:2016qru, Bastero-Gil:2019gao, Rosa:2019jci, Rosa:2018iff, Levy:2020zfo}, where the mass of the produced particles is an oscillatory (and therefore bounded) function of the inflaton field, so that they may remain light throughout inflation even if the inflaton attains very large (even super-planckian) values. An additional discrete interchange symmetry is also imposed to cancel the leading contributions to the inflaton's thermal mass of the two particle species it couples to, although it has recently been shown \cite{Ferraz:2023qia} that the thermal backreaction is under control even for a single species and without imposing any symmetry\footnote{For other proposals to realize warm inflation with axion-like fields see \cite{Berghaus:2019whh,Zell:2024vfn,Berghaus:2025dqi,Ito:2025lcg,Broadberry:2025ggb}.}.

In this work, we show that the same particle physics setup behind the Warm Little Inflaton scenario can lead to a {\it non-thermal} particle production mechanism during the slow-roll phase with similar features to conventional scalar preheating scenarios. This therefore alleviates the constraints posed by particle production close to thermal equilibrium.

 Anticipating our main results, we will show that efficient particle production of scalars due to a (narrow) parametric resonance is allowed to occur during slow-roll. This is achieved within a quantum field theory where the inflaton is coupled to a single additional scalar degree of freedom $\chi$, similarly to Ref. \cite{Ferraz:2023qia}, yielding an oscillatory mass term for this field. We will show that this process introduces small modifications in both the curvature and the tensor power spectra that may fall within the range of future CMB observations.

	This work is organised as follows. In the next section, we introduce our particle physics model and compute the comoving number density and the physical energy density of the produced $\chi$ quanta. In \Cref{sec:backreaction}, we compute the backreaction of these particles on the inflaton field: the effect on the classical inflaton is discussed in \Cref{subsec:classical}, while the effect on its quantum fluctuations is discussed in \Cref{subsec:quantum}. In each subsection, we analyse the impact of the backreaction on several CMB observables. In \Cref{sec:gravitational_waves}, we determine the gravitational wave spectrum induced by the $\chi$ quanta, in particular its contribution to the primordial tensor power spectrum.~Our theoretical results are tested against observational data for two illustrative examples of inflationary potentials, and the allowed parameter space of our model is determined for each case. In \Cref{sec:discussion_conclusion}, we summarise and discuss our results, and anticipate some possible future work.
	
	 We use the metric signature $(+,-,-,-)$ and consider natural units $\hbar = c = 1$ and the reduced Planck mass $M_\text{P}^{-2} = 8 \pi G$ throughout our discussion.

	\section{Resonant particle production during inflation}
	\label{sec:resonant_PP}
	
	In our setup, the inflaton field appears as a scalar degree of freedom arising from the collective spontaneous breaking of a U(1) gauge symmetry, under which it is invariant. As in Refs. \cite{Ferraz:2023qia, Bastero-Gil:2019gao}, we consider two complex scalar fields $\Phi_1$ and $\Phi_2$, equally charged under the U(1) symmetry, and each subject to a Higgs-like potential, so that they both attain the same non-zero vacuum expectation value (VEV) $\braket{\Phi_1} = \braket{\Phi_2} = \frac{M}{\sqrt{2}}$. These fields can then be parametrised as
	\begin{equation}\label{eq:field_parametrisation}
			\Phi_1 = \frac{M+h_1}{\sqrt{2}}\, e^{i \frac{ \theta+\phi}{M}}\:\:, \qquad	\Phi_2 = \frac{M+h_2}{\sqrt{2}}\, e^{i \frac{\theta-\phi}{M}} ~,
	\end{equation}
	where $h_{1,2}(x)$ are Higgs-like fields and $\theta(x)$ is the Nambu-Goldstone boson (NG) of the spontaneously broken U(1) symmetry that can be rotated away in the unitary gauge, effectively being absorbed by the now massive U(1) gauge boson. The relative phase $\phi(x)$ corresponds to a physical scalar degree of freedom that we identify with the inflaton field, which is indeed a singlet of the U(1) gauge symmetry\footnote{In the literature, the Warm Little Inflaton has been referred to as a pseudo-NG boson, although this may be misleading since it is a gauge singlet and its scalar potential is not constrained by a shift symmetry.}. 
	 The full Lagrangian density of our model is thus
\begin{equation} \label{eq:fulllagrangian}
	\begin{split}
		\mathcal{L} \:=\:& \left( D_\mu \Phi_1\right)^* \left( D^\mu \Phi_1\right) \,+\, \left( D_\mu \Phi_2\right)^* \left( D^\mu \Phi_2\right) \,+\, \frac{1}{2}\partial_\mu \chi \, \partial^\mu \chi \,-\, \frac{1}{4}F_{\mu \nu}F^{\mu \nu}  \\
		&-\, \sum_{i=1,2}\frac{\lambda_i}{4}\left( \vert \Phi_i \vert^2 - \frac{M^2}{2} \right)^2 -\, V(\phi)
		- \frac{1}{2}\, g^2 \vert  \Phi_1 - \Phi_2 \vert^2  \chi^2 ~,
	\end{split}
\end{equation} 
	\noindent where, as permitted by the U(1) symmetry, we also include a potential term for $\phi(x)$. We have also introduced a single additional U(1)-neutral, real scalar field $\chi$, which describes the particles produced during inflation. We note that any coupling of the form $-\, \frac{1}{2}\, g^2 \vert  a\Phi_1 \,+\,b \Phi_2 \vert^2  \chi^2$, $a,b \in \mathbb{C}\setminus \{0\}$, would yield analogous results; here, we focus on the simple case $a=-b=1$.  The mass scale $M$ and the dimensionless coupling constant $g<1$ are the free parameters of the model. In the unitary gauge, the last term in \eqref{eq:fulllagrangian} yields a squared mass $m_\chi^2(\phi)= 2g^2M^2 \sin^2 \big( \frac{\phi}{M} \big)$ for the $\chi$ field, which is as anticipated an oscillatory function of the inflaton field\footnote{We discard higher-order interaction terms since we are mainly concerned with analysing $\chi$ production. We therefore
ignore the $h_{1,2}$ and $A_\mu$ degrees of freedom, which have constant masses $\sim M \gg H$ during inflation.}.

Let us begin by analyzing the production of $\chi$-particles due to the motion of the classical, homogeneous, inflaton field, ignoring backreaction, which we will compute later in our discussion. Hence, since $\phi(t)$ is in general a monotonic function in the slow-roll phase, the mass of the $\chi$-particles will oscillate in time, akin to what happens in the conventional preheating stage, although in the latter the mass is typically linear in the field and it is the field that oscillates in time.

Henceforth, we consider a flat FLRW universe dominated by the slowly-rolling inflaton field $\phi(t)$, and take the Hubble rate $H$ to be approximately constant, which we shall see is a good approximation in the context of $\chi$ production. The equation of motion and energy density of the $\chi$ field are therefore given by
	\begin{equation}\label{eq:chiEoM}
		\ddot \chi + 3H \dot \chi - \frac{1}{a^2} \nabla^2 \chi + m_\chi^2(t)\, \chi = 0~,\quad		\rho_\chi \equiv T_\chi^{00} =  \frac{1}{2}  \dot \chi ^2 + \frac{1}{2} \frac{\vert \nabla \chi \vert^2}{a^2} +  \frac{1}{2} m_\chi^2(t)\,  \chi ^2 ~ ,
	\end{equation}
	with the $\chi$ field then being promoted to a quantum operator $\hat \chi$ and expressed as a sum of Fourier modes. Defining $\omega_k^2(t) = \frac{k^2}{a^2(t)} + m_\chi^2(t)$ to be the frequency of each mode, we then have in Fourier space
	\begin{equation}\label{eq:chimodesDHO}
		\ddot \chi_k + 3H \dot \chi_k + \omega_k^2(t) \,\chi_k = 0~,
	\end{equation}
	which greatly resembles the equation of motion for a damped harmonic oscillator with a time-varying frequency, where the damping is due to Hubble expansion. Since during slow-roll the time derivatives of $\phi$ get successively smaller as their order increases, we may perform a Taylor expansion of $\phi$ around some instant $t_0$ and keep only the first two terms $\phi(t) \simeq \phi(t_0) + \dot \phi(t_0)(t-t_0) \equiv \dot \phi t + \delta$, yielding $m_\chi^2(t)\simeq 2g^2M^2 \sin^2 \big( \frac{\dot \phi t}{M}\big)$, where we have set $\delta = 0$ since a phase factor does not alter the results. Defining the function $X_k = a^{3/2} \chi_k$ and switching to the rescaled time variable $z = \frac{|\dot \phi|}{M}\,t$, we have
	\begin{equation}\label{eq:chimodesMathieu}
		X_k'' + \left[ A_k(z) - 2q \cos( 2z ) \right] X_k = 0~,\qquad 		\begin{cases}
			A_k(z) = \left(\frac{M}{\dot \phi}\right)^2 \left(\frac{k^2}{a^2(z)} - \frac{9}{4}H^2\right) + 2q \\
			q = \frac{1}{2}\left(\frac{M}{\dot \phi} \right)^2 g^2M^2
		\end{cases}~,
	\end{equation}
where the primes denote derivatives with respect to $z$.

Equation \eqref{eq:chimodesMathieu} is a Mathieu-like equation with a variable parameter $A_k(z)$ \cite{mclachlan1947theory,Kofman:1997yn, NIST}. Rigorously, both $A_k$ and $q$ vary in time; however, we may neglect the relative variation of $\dot \phi$ during slow-roll when compared with that of the scale factor, and thus take $q$ to be constant. In fact, the theory of Mathieu equations assumes both $A_k$ and $q$ to be constant, but we may safely apply it to this case as long as we ensure that the variation of both parameters is adiabatic \cite{Rosa:2007dr}. The relative variation of $q$ due to the slow-roll dynamics can be shown to be given by $\frac{q'}{q} = 2 \left( \eta_V - \epsilon_V \right)$ in terms of the potential slow-roll parameters, which is indeed small. The variation of $A_k(z)$ is trickier to analyse; nonetheless, we shall see below that under some conditions it is also adiabatic.

 The main feature of Mathieu equations is the existence of parametric resonance bands for some values of the parameters $A_k$ and $q$. We will only consider the case of a narrow resonance ($q \ll 1$), for which the resonance bands occur for $A_k \sim n^2$, with $n$ a positive integer, given that in this case particle production is more moderate than for a broad resonance ($q \gtrsim 1$), and that the inflaton field should remain as the dominant component. We shall also restrict our analysis to the first ($A_k \sim 1$) and most important of these bands, which occurs for $1-q \lesssim A_k \lesssim 1+q$ \cite{mclachlan1947theory,Kofman:1997yn}.

Let us briefly focus on the parameter $A_k(z)$. Since we expect the modes being produced (i.e. the ones inside the resonance band) to be causally connected, we must impose that for those modes $\frac{k}{a(z)} \gg H$, so that, at the centre of the resonance band (i.e.~for $A_k \simeq 1$), we have $\big(\frac{k/a}{gM}\big)^2 \simeq \frac{1}{2q} - 1  \gg 1$. Since $\langle m_\chi \rangle \equiv gM$ is the average value of the $\chi$-particles mass, this implies that the produced particles are relativistic. Furthermore, this allows us to write $A_k(z) \simeq \frac{1}{2\epsilon_V} \left( \frac{k}{a(z)H} \right)^2 \left( \frac{M}{M_\text{P}} \right)^2 $, where $\dot \phi^2 = 2 \epsilon_V M_\text{P}^2 H^2$ was used.

Going back to our main discussion, Floquet's theorem \cite{mclachlan1947theory,NIST} states that, for $q \ll 1$ and $A_k(z) \sim 1$, the solution of the Mathieu equation evolves as $X_k(z) \propto e^{\mu_k(z)\, z}$, where $\mu_k(z)$ is the Floquet exponent and is given by
\begin{equation}\label{eq:chimodesFloquet}
	\mu_k(z) = \frac{1}{2} \sqrt{q^2 - [A_k(z)-1]^2}~,
\end{equation}
which attains its maximum value $\mu_k^\text{max} = \frac{q}{2}$ at $A_k(z)=1$.
Inside the first resonance band, $\mu_k(z)$ is real (and positive), such that a mode with comoving momentum $k$ enters the band when $A_k(z_+) = 1+q$ and exits it once $A_k(z_-)=1-q$ (notice that $A_k(z)$ decreases as inflation progresses). We note that, at any given time, the mode that is at the centre of the resonance band has a physical momentum $p_c \equiv \left(\frac{k}{a}\right)_c = \frac{gM}{\sqrt{2q}}$, which is a quantity that only varies adiabatically due to $q$. Moreover, the physical momenta of \emph{all} modes that are inside the resonance band at a certain time are in fact very close to $p_c$, so that the momentum distribution of the $\chi$-particles being produced should resemble a Dirac peak centred at $p_c$. Reverting momentarily to cosmic time $t$, it is also simple to show, using $A_k(t_\pm) = 1\pm q$  that $t_\pm(k) \simeq t_c(k) \mp \frac{q}{2H} $, where $t_c(k) = H^{-1} \ln \frac{\sqrt{2q}\: k}{gM \:a_0}$ is the time at which the mode $k$ is at the centre of the band, so that each mode spends a time $\Delta t = qH^{-1}$ inside the resonance band , i.e.~$\Delta N_e =  q\ll 1$ $e$-folds. This result further validates the constant-$H$ approximation we are considering: since each $k$ mode spends much less than one $e$-fold in the resonance band, during that period $H$ is in fact constant to a very good approximation. Surely, throughout inflation, different modes experience different values of $H$ during their passage through the resonance band, but for each individual mode that value is essentially the same for the entire passage.

For the modes inside the resonance band, we may Taylor expand the scale factor $a(t)$ around $t_c(k)$ and keep only the first-order term. This in turn results in an $A_k(z)$ parameter given by $A_k(z) \simeq 1 - \gamma[z-z_c(k)]$, where $\gamma = \sqrt{\frac{2}{\epsilon_V}} \frac{M}{M_\text{P}}$ must ideally be a small quantity in order for the variation of $A_k(z)$ to be adiabatic. Indeed, we have found that $\gamma \ll 1$ in the interesting region of the parameter space $(g,M)$ for all considered inflationary models, so that adiabaticity is ensured. Notice that, since $\epsilon_V \lesssim 1$, this condition implies that $M \ll M_\text{P}$. Moreover, it can be shown that, like $q$, $\gamma$ varies adiabatically, cf. $\frac{\gamma'}{\gamma} = \eta_V - 2\epsilon_V$.

Rewriting $p_c$ in terms of $\gamma$ yields $p_c = \frac{2H}{\gamma}$, at which point we note that the frequency at which $m_\chi(t)$ oscillates is essentially equal to the frequency associated with the particles being produced, since for those particles $\omega_k(t)  \simeq \left(\frac{k}{a}\right)_c = \frac{2H}{\gamma} = \frac{|\dot \phi|}{M}$. This result is in line with other common resonant systems, where the resonances arise precisely from the proximity or equality between two different frequencies, one inherent to the system ($\omega_k$) and one imposed on it ($\frac{|\dot \phi|}{M}$). 

The condition that the $k$ modes be sub-horizon while inside the resonance band implies that $m_\chi(t)$ must complete at least one oscillation within an $e$-fold of inflation, which is required in order to have efficient particle production. We must then also ensure that the $\chi$ mass completes at least one oscillation while each mode is inside the resonance band, i.e. $\frac{|\dot \phi|}{M} \, \Delta t >  2\pi$, or equivalently $\frac{q}{\gamma} > \pi$, which we have found to hold in a significant region of the allowed parameter space for all inflationary models considered in this work. Furthermore, we consider only scenarios in which $\langle m_\chi \rangle >H$, so that we may neglect $\mathcal{O}(H)$ quantum corrections to $m_\chi$ \cite{Baumann:2014nda}, as well as sub-Planckian particle masses.

	As mentioned before, resonant production of particles occurs while a mode is inside the resonance band. We saw in the previous section that the theory of Mathieu equations \cite{Kofman:1997yn, mclachlan1947theory, NIST} can be approximately applied to Eq.~\eqref{eq:chimodesMathieu}, so that, from Floquet's theorem, the solution to this equation can be written approximately as the product of a periodic function $P_k(z)$ by an exponential $e^{\mu_k(z) \,z}$, which we can approximate as $X_k(z) = P_k(z) \,e^{\mu_k(z) z} \simeq (2 \omega_k^0)^{-1/2}\, e^{\pm i \omega_k \frac{\gamma}{2H} z}\,e^{\mu_k(z) \,z}$, where the normalization is chosen so as to ensure a vanishing occupation number for each $k$ mode before it entered  the resonance band, with $\omega_k^0 \sim \frac{2H}{\gamma}$ denoting the corresponding frequency.
	
	 The comoving energy density and comoving occupation number are then given by \cite{Kofman:1997yn}
	\begin{equation}\label{eq:comoving_number}
	\tilde \rho_k \:\:\equiv\:\: \frac{1}{2} \left( \vert \dot X_k \vert^2 + \omega_k^2\, \vert X_k \vert^2 \right)~,\quad \tilde n_k \:\:\equiv\:\: \frac{\tilde \rho_k}{\omega_k} - \frac{1}{2} \:\:\simeq\:\: \frac{1}{2} \frac{\omega_k}{\omega^0_k} \,e^{2 \mu_k(z) \,z} - \frac{1}{2} ~,
	\end{equation}
the former being related to the vacuum expectation value of the operator $\hat \rho_\chi$ via $\langle 0|\hat \rho_\chi |0 \rangle = \frac{1}{a^3}\int \frac{d^3 k}{(2 \pi)^3}\: \tilde \rho_k \equiv \rho_\chi$, and where we assumed that $\dot X_k(t) \simeq \pm\, i \omega_k X_k(t)$. Moreover, if a mode $k$ is inside the resonance band, in an interval $dz$ the function $X_k$ will be amplified by a factor $e^{\mu_k(z)\,dz}$, so that between two instants $z_i$ and $z_f$ the total amplification of $X_k$ is $\sim e^{\mu_k z}$, where 
	\begin{equation}\label{eq:integratedFloquetexponent}
		\begin{split}
	\mu_k z\, \equiv\, \int_{z_i}^{z_f} \mu_k(z)\: dz \,=\, \frac{q^2}{4\gamma} \Biggl\{ &\arcsin\left[ \frac{\gamma}{q}(z_f-z_c) \right] + \frac{\gamma}{q}(z_f-z_c) \sqrt{1-\left(\frac{\gamma}{q}\right)^2(z_f-z_c)^2} \\
	&- \arcsin\left[ \frac{\gamma}{q}(z_i-z_c) \right] - \frac{\gamma}{q}(z_i-z_c) \sqrt{1-\left(\frac{\gamma}{q}\right)^2(z_i-z_c)^2}\Biggr\}~,
		\end{split}
	\end{equation}
making $X_k(z_i \rightarrow z_f) \simeq (2 \omega_k^0)^{-1/2} \,e^{\pm i \omega_k \frac{\gamma}{2H} z}\,e^{\mu_k z}$ \cite{Rosa:2007dr}. The above quantity attains its maximum value of $\frac{\pi q^2}{4 \gamma}$ when $z_i = z_+(k)$ and $z_f=z_-(k)$, corresponding to a mode $k$ that has been through the entire resonance band. Hence, from \eqref{eq:comoving_number}, the comoving occupation number and the total comoving number density of $k$-momentum particles produced between $z_i$ and $z_f$ are respectively given by
	\begin{equation}\label{eq:integrated_comoving_number}
		\tilde n_k(z_i \rightarrow z_f) \,\simeq\, \frac{1}{2} \frac{\omega_k}{\omega^0_k}\, e^{2\mu_k z} - \frac{1}{2} ~,\quad \tilde n_\chi (z_i \rightarrow z_f) \,= \,\int \frac{d^3k}{(2\pi)^3}\,\: \tilde n_k(z_i \rightarrow z_f)~.
	\end{equation}
	Notice that the comoving occupation number for a certain $k$ mode varies in time approximately as a step function centred in $t_c(k)$ \cite{Rosa:2007dr}; moreover, it should be noted that the value of $\tilde n_k(z_i \rightarrow z_f)$ obtained using the approximate analytical expression in Eq.~\eqref{eq:integratedFloquetexponent} slightly underestimates the analogous value obtained from numerically solving the Mathieu equation. Notice also that when the expression for $\tilde n_k(z_i \rightarrow z_f)$ is inserted into the 3-momentum integral a $\frac{1}{2}k^2$ term appears in the integrand function, causing the integral to diverge. This behaviour is related to the energy content of the vacuum and should be dealt with through renormalisation. However, given the fact that we are concerned only with the number of $\chi$-particles that are produced during inflation, and so with the energy that is added to that of the vacuum by said particles, we may neglect that part of the integral altogether, which is what we will do henceforth, the expression for the integrand function then simply becoming $\frac{1}{2}k^2\frac{\omega_k}{\omega^0_k}\,e^{2\mu_k z}$. Likewise, the total physical energy density of particles produced between $z_i$ and $z_f$ is, at a time $z > z_f$, given by
		\begin{equation}\label{eq:totalphysicalenergydensity}
		\rho_\chi (z_i \rightarrow z_f) \:\simeq \: \frac{1}{a^3(z)}\int \frac{d^3k}{(2\pi)^3}\,\: \omega_k\left(\tilde n_k(z_i \rightarrow z_f) + \frac{1}{2}\right) ~.
	\end{equation}
	
At any given time $z$, the 3-momentum integrals are dominated by the modes within the resonance band, and a good approximation to the total particle number can be obtained by integrating up to the mode entering the resonance band $k_\text{RB}(z) = \frac{2H}{\gamma}\,a(z)\,e^{-\frac{q}{2}}$. The lower integration limit is irrelevant, since modes with $k\ll k_\text{RB}(z)$ have already been diluted by expansion after crossing the resonance band at earlier times. This yields for the total number density and energy density of the produced $\chi$-particles
			\begin{equation}\label{eq:final_results_n_rho}
		n_\chi\simeq  \frac{2H^3}{3\pi^2 \gamma^3} e^{\xi- \frac{3}{2}q} ~,\quad \rho_\chi  \simeq   \frac{H^4}{\pi^2 \gamma^4} \,e^{\xi - 2q}~,
	\end{equation}
where $\xi \equiv \frac{\pi q^2}{2 \gamma}$ thus quantifies the resonance efficiency. These are approximately constant, up to slow-roll evolution of the parameters. In particular, $\frac{\xi'}{\xi} = 3\eta_V - 2\epsilon_V$, yielding $\frac{\rho_\chi'}{\rho_\chi} = 4(\epsilon_V - \eta_V) + 4q\,(\epsilon_V - \eta_V) + \xi \,(3 \eta_V - 2\epsilon_V)$. This is due to a balance between the $\chi$-particle production rate and their Hubble dilution rate. In fact,  there is always a $k$-mode inside the resonance band, the production of which nearly compensates the dilution of the particles produced at earlier times.

This thus yields a novel way of sustaining a bath of relativistic particles during inflation, albeit with a {\it non-thermal} distribution peaked around the physical momentum $p_c\simeq 2H/\gamma$. Curiously, for $\gamma\ll 1$, we may have $\rho_\chi \gg H^4$ even if individual modes are not exponentially amplified, simply due to the large density of comoving momentum modes within the resonance band at any given time.

\begin{figure}[htbp]
	\centering
	\begin{subfigure}{0.489\linewidth}
		\centering
		\includegraphics[width=1\linewidth]{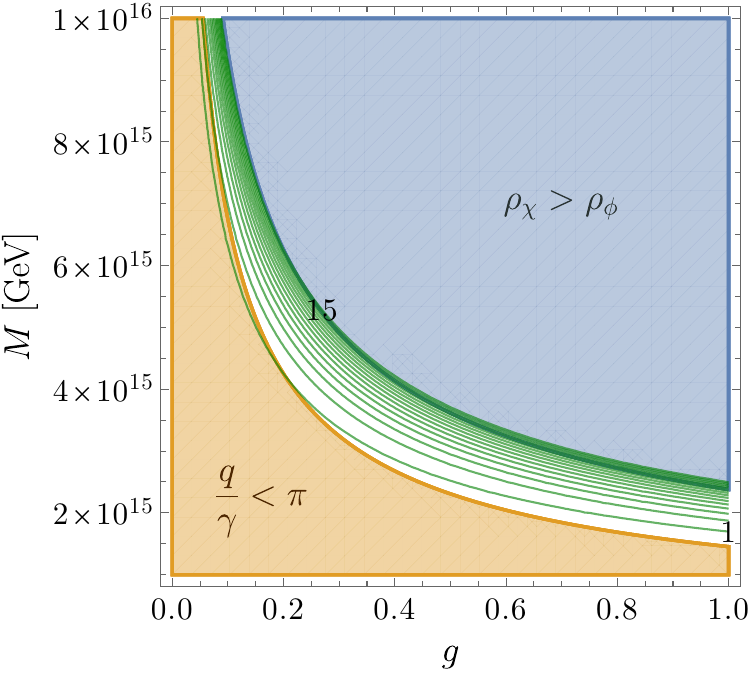}
		\caption{}
		\label{fig:regionplot_quad_monomial}
	\end{subfigure}\hspace*{8pt}
	\begin{subfigure}{0.489\linewidth}
		\centering
		\includegraphics[width=1\linewidth]{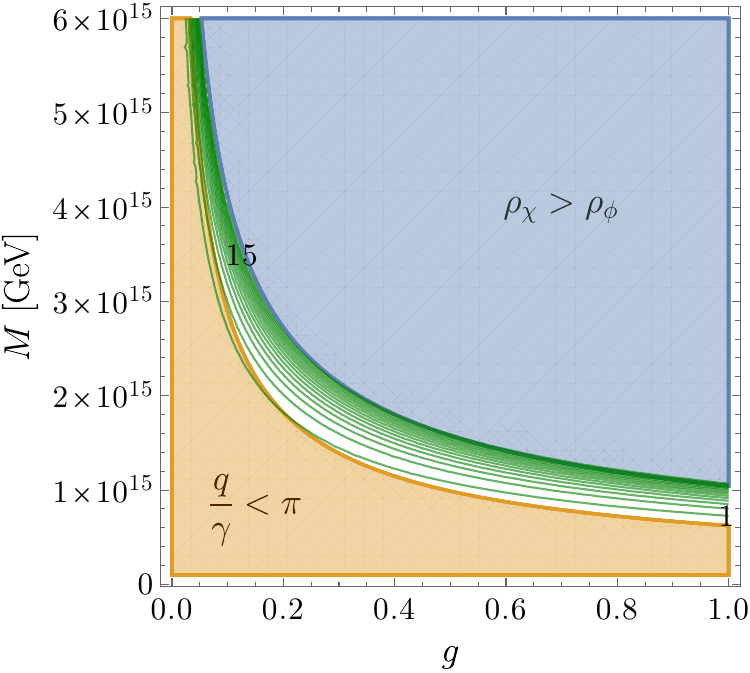}
		\caption{}
		\label{fig:regionplot_quad_hilltop}
	\end{subfigure}
	\caption{Parameter space $(g,M)$ for (a) the quadratic monomial potential with $\phi_i = 15\, M_\text{P}$ (at $\epsilon_V \simeq 0.009$), and (b) the quadratic hilltop potential with $\phi_i = 5\, M_\text{P}$ and $\kappa =  10^{-2}$ (at $\epsilon_V \simeq 0.002$). The allowed region, not excluded by the conditions $\frac{q}{\gamma}< \pi$ (orange) and $\rho_\chi > \rho_\phi$ (blue), is shown in white (additional conditions are non-restrictive). Contour lines for the resonance parameter $\xi$ (for $1 \leqslant \xi \leqslant 15$) are shown in green.}
	\label{fig:regionplots}
\end{figure}

In Figures \ref{fig:regionplot_quad_monomial} and \ref{fig:regionplot_quad_hilltop} we show the parametric regions in the $(g,M)$ plane where $\chi$-particle production can consistently occur without exceeding the inflaton's energy density, for a quadratic monomial potential\footnote{Although monomial potentials have been ruled out by observations \cite{Planck:2018jri,ACT:2025tim}, we consider these models for their simplicity and historical relevance.}  $V(\phi) = \frac{1}{2}m^2 \phi^2$ and for a quadratic hilltop potential $V(\phi) = V_0 \big[1- \frac{\kappa}{2} \big(\frac{\phi}{M_\text{P}}\big)^2\big]$, respectively. We also show in each case the contours of constant resonance efficiency parameter $\xi$. All quantities are evaluated 60 $e$-folds before the end of inflation, i.e.~when the relevant CMB scales crossed the horizon.~As one can easily see in this figure, the produced particles are quite heavy, with masses around the grand unification scale, $gM\gtrsim 10^{15}$ GeV, even though they are relativistic.

To better assess the impact of the parametric resonance on the dynamics of inflation, we discuss in the next section the leading backreaction effects.

	\section{Backreaction on the inflationary dynamics}
	\label{sec:backreaction}
	
	We shall split the effect of the backreaction into two parts: the backreaction on the classical, homogeneous inflaton, and the backreaction on the quantum fluctuations of the inflaton. These will be dealt with independently, since any interference between the two corresponds to a higher-order correction.
	
	\subsection{Background dynamics}
	\label{subsec:classical}
	
	Recovering the Lagrangian from Eq.~\eqref{eq:fulllagrangian}, we may define the following Lagrangian
	\begin{equation} \label{eq:backreactionlagrangian}
		\mathcal{L}_{\phi \chi} \:=\:  \frac{1}{2} \partial_\mu \phi \, \partial^\mu \phi  + \frac{1}{2} \partial_\mu \chi \, \partial^\mu \chi - \frac{1}{2} m_\chi^2(\phi) \, \chi^2 - V(\phi)~,
	\end{equation}
	where, as before, $m_\chi^2(\phi) = 2\,g^2 M^2 \sin^2 \big( \frac{\phi}{M} \big)$, allowing us to define an action functional $S[\phi, \chi] \,= \int d^4x \,\sqrt{-g} \:\mathcal{L}_{\phi \chi}$.
	It can be shown using an effective action formalism that the effective EoM for the inflaton field receives two contributions: a Coleman-Weinberg (CW) term \cite{Coleman:1973jx,Armendariz-Picon:2020tkc,Bastero-Gil:2016qru} and a term proportional to $\tilde n_k$ \cite{Kofman:1997yn}, both of which can be written as a trace in momentum space. The CW term can be integrated and renormalised using the $\overline{\text{MS}}$ scheme \cite{Bastero-Gil:2016qru}, leading to a contribution to the effective EoM given by 
	\begin{equation} \label{eq:CW_EoM_cont_renormalised}
		\Delta V'_\text{CW}(\phi) \:\equiv\: \frac{g^4 M^3}{8\pi^2}\,\sin^2 \left(\frac{\phi}{M}\right) \sin \left(\frac{2\phi}{M}\right) \ln \left(\frac{\mu^2}{m_\chi^2}\right) ~,
	\end{equation}
where $\mu$ is the $\overline{\text{MS}}$ renormalisation scale. We immediately see that if we select $\mu = m_\chi$ this contribution vanishes altogether. Nonetheless, let us compare it with the one proportional to $\tilde n_k$. For this, we anticipate an upcoming result (cf. Eq.~\eqref{eq:BRfinalEoM} below): that the remaining contribution to the EoM is given by
	\begin{equation} \label{eq:PP_EoM_cont_renormalised}
		\Delta V'_\text{PP}(\phi) \:\equiv\: \frac{g^2 M}{2 \pi^2}  \left(\frac{H}{ \gamma}\right)^2 e^{\xi-q} \:\sin\left(\frac{2  \phi}{M}\right) ~,
	\end{equation}
where the subscript indicates that this quantity is related to $\chi$-particle production. In spite of the clear suppression due to an additional $g^2$ factor in Eq.~\eqref{eq:CW_EoM_cont_renormalised}, these two contributions are not straightforward to compare using only their analytical expressions. Numerically, however, we found that for a renormalisation scale $H < \mu < M_\text{P}$ the CW term is subleading relative to $\Delta V'_\text{PP}$ for all inflationary models we considered in this work; hence, we may neglect the former in our computations and consider only the contribution coming from particle production.

In fact, this amounts to using the Hartree approximation\footnote{In Ref.~\cite{Kofman:1997yn}, some additional contributions are shown to be subdominant relative to the Hartree approximation. In the present work, we will not perform such an analysis and will consider exclusively this approximation, which will in principle be able to capture the main behaviour of the backreaction; however, a more rigorous study of this system is certainly of interest.} for the backreaction on the inflaton field \cite{Kofman:1997yn,Armendariz-Picon:2020tkc}.~Neglecting for now the quantum fluctuations of the inflaton, the EoM for $\phi$ can be obtained simply by replacing $\chi^2$ with its vacuum expectation value $\langle \chi^2 \rangle  =  \int \frac{d^3 k}{(2 \pi)^3}\: \vert \chi_k \vert^2$ in the Lagrangian \eqref{eq:backreactionlagrangian} and varying $S[\phi,\chi]$ with respect to $\phi$
	\begin{equation} \label{eq:hartreeEoM}
		\Box  \phi +  V' + \frac{1}{2} ( m_\chi^2)' \, \langle \chi^2 \rangle = 0~,
	\end{equation}
where $ V \equiv V( \phi)$ and $ m_\chi^2 \equiv m_\chi^2( \phi)$, with the primes denoting derivatives with respect to $ \phi$. At this point, we may note that the Hartree contribution to the backreaction may be entirely removed if instead of considering a single scalar field $\chi$ whose quanta are produced via parametric resonance we consider two scalar fields $\chi_1$ and $\chi_2$, with interaction terms  $-\frac{1}{2} g^2 \vert  \Phi_1 \,+\, \Phi_2 \vert^2  \chi_1^2 - \frac{1}{2} g^2 \vert  \Phi_1 \,-\, \Phi_2 \vert^2  \chi_2^2$, invariant under the discrete interchange symmetry $\Phi_1 \leftrightarrow i\Phi_2$ and $\chi_1 \leftrightarrow \chi_2$ \cite{Bastero-Gil:2019gao,Bastero-Gil:2016qru}. The effective oscillating masses for each field $\chi_{1,2}$ will then be $m_{\chi_1} \propto \sin\big(\frac{\phi}{M}\big)$ and $m_{\chi_2} \propto \cos\big(\frac{\phi}{M}\big)$. However, the quanta of each field $\chi_{1,2}$ are produced via the same process as described for $\chi$, leading to variances $\langle \chi_1^2 \rangle = \langle \chi_2^2 \rangle = \langle \chi^2 \rangle$, which means that in the Hartree approximation the contributions of each field $\chi_{1,2}$ would cancel each other, cf. Eq.~\eqref{eq:hartreeEoM}, leaving us only with the subleading CW term (notice that this says nothing about the backreaction beyond the Hartree approximation, which may still contribute as well). This is, in fact, completely analogous to the cancellation of the leading thermal corrections to the inflaton's mass in the Warm Little Inflaton model, even if we are considering a non-thermal particle production mechanism.
	
	Returning to our discussion, in order to compute $\langle \chi^2 \rangle$, we write the solution of equation \eqref{eq:chimodesDHO} in terms of Bogoliubov coefficients $\alpha_k = \alpha_k(t)$ and $\beta_k = \beta_k(t)$ \cite{Kofman:1997yn,Kachelriess:2022jmh}, yielding
	\begin{equation} \label{eq:expvalchibogoliubov}
			\vert \chi_k \vert^2 \,=\, \frac{a^{-3}}{2 \omega_k}\,\left[ \vert \alpha_k \vert^2  + \vert \beta_k \vert^2 + 2\, \mathfrak{Re} \left( \alpha_k\, \beta_k^* \,\, e^{-2i \int \omega_k \, dt} \right) \right] \,\simeq\, \frac{a^{-3}}{2 \omega_k}\,\left( 1  + 2\, \vert \beta_k \vert^2 \right) ~,
	\end{equation}
 where we dropped the high-frequency term (as it should provide a subdominant contribution \cite{Kofman:1997yn}). Moreover, since $\vert \beta_k \vert^2 = \tilde n_k$, we have
	\begin{equation} \label{eq:expvalfinal}
		\langle \chi^2 \rangle  \:\simeq\: \frac{1}{a^3} \int \frac{d^3 k}{(2 \pi)^3} \, \frac{\tilde n_k + \frac{1}{2}}{\omega_k} \:\simeq\: \frac{1}{2 \pi^2\,a^3} \int_{0}^{k_\text{RB}(z)} k^2\,dk \:\: \frac{\frac{1}{2}\,e^\xi}{k/a}~,
	\end{equation}
where once again the major contribution to the integral comes from momenta that have crossed the entirety of the resonance band by a time $z = \frac{2H}{\gamma}t$, and where we used the fact that $\omega_k^0 \simeq \omega_k \simeq \frac{k}{a}$ for the modes within the narrow resonance band that dominate the particle production spectrum. This leads to an effective EoM for $ \phi$ that reads
	\begin{equation} \label{eq:BRfinalEoM}
		\Box  \phi +  V' + \frac{g^2 M}{2 \pi^2}  \left(\frac{H}{ \gamma}\right)^2 e^{\xi-q} \:\sin\left(\frac{2  \phi}{M}\right) = 0~,
	\end{equation}
in accordance with the result we anticipated in Eq.~\eqref{eq:PP_EoM_cont_renormalised}, and which we may use to define an effective inflaton potential	$\mathcal{V}( \phi) \equiv V( \phi) \,+\,\Lambda^4 \, \cos \big(\frac{2 \phi}{M}\big)$, where $\Lambda^4 \equiv - \frac{g^2 M^2}{4 \pi^2}  \big(\frac{H}{ \gamma}\big)^2 e^{\xi-q}$ varies adiabatically due to the slow-roll dynamics, and so is taken to be approximately constant throughout inflation.

 In addition, we may define a pair of effective Hubble slow-roll parameters, using an effective Hubble parameter $\mathcal{H}$, defined by the Friedmann equation $\mathcal{V}( \phi) \equiv 3 M_\text{P}^2 \mathcal{H}^2 $,
\begin{equation} \label{eq:QCslowrollparamshubble}
	\epsilon_\mathcal{H} \equiv -\frac{ \mathcal{\dot H}}{\mathcal{H}^2} \,,\quad \eta_\mathcal{H} \equiv  2\,\epsilon_\mathcal{H} - \frac{1}{2}\, \frac{\dot \epsilon_\mathcal{H}}{\epsilon_\mathcal{H} \, \mathcal{H}}~.
\end{equation}

	We may now try to find an analytical solution to Eq.~\eqref{eq:BRfinalEoM}, now rewritten as $\Box  \phi + \mathcal{V}'( \phi) = 0$, where the d'Alembertian is now defined using $\mathcal{H}$, rather than $H$. This allows us to follow Ref. \cite{Flauger:2010ja} somewhat closely. As such, we expand the field as $ \phi = \phi_0 + \phi_1 + ...$, where $\phi_0$ is the uncorrected, homogeneous inflaton, and $\phi_1$ is the first-order correction (linear in $\Lambda^4$) due to the backreaction of $\chi$. Hence, the slow-roll Friedmann equation and the slow-roll Klein-Gordon equation for $\phi_0$ yield $\dot \phi_0 = - M_\text{P}\, \frac{V_{,\phi_0}(\phi_0)}{\sqrt{3\,V(\phi_0)}}$, which can of course be solved independently of $\phi_1$. With this, we only need to find the EoM for $\phi_1$ alone. Expanding $\mathcal{V}( \phi)$ and $\mathcal{H}(\phi)$ to first-order in $\phi_1$ and inserting the results into $\Box  \phi + \mathcal{V}'( \phi) = 0$ eventually leads to (keeping only linear terms in $\phi_1$)
	\begin{equation} \label{eq:EoM_phi1_phi_0}
		\phi_1'' \:-\: \frac{3}{M_\text{P}}\, \frac{1}{\sqrt{2 \epsilon_{V_0}}}\, \phi_1' \:+\: \frac{3}{2M_\text{P}^2} \, \left(\frac{\eta_{V_0}}{\epsilon_{V_0}} - 1\right)\,\phi_1 \: =\: \frac{3\,\Lambda^4}{ \epsilon_{V_0} V_0 M}\,\sin\left(\frac{2 \phi_0}{M}\right)~,
	\end{equation}
	where $\epsilon_{V_0}$ and $\eta_{V_0}$ are computed using the potential $V_0 \equiv V(\phi_0) \equiv 3 M_\text{P}^2 H_0^2$, and where the primes now denote derivatives with respect to $\phi_0$, given that in order to arrive at this equation, we converted the cosmic time derivatives of $\phi_1$ into derivatives with respect to $\phi_0$, using $\dot \phi_1 = \frac{d \phi_1 }{d\phi_0}\, \dot \phi_0$ and the EoM for $\phi_0$. We have furthermore assumed that $|\phi_1| \ll \frac{M}{2}$ (which we shall see is the case), and used the fact that $\big|\Lambda^4\big| \ll \big|\frac{M}{2}\,V_{,\phi_0}(\phi_0)\big|$.

\begin{figure}[htbp]
		\centering
		\begin{subfigure}{0.48\linewidth}
			\centering
			\includegraphics[width=1\linewidth]{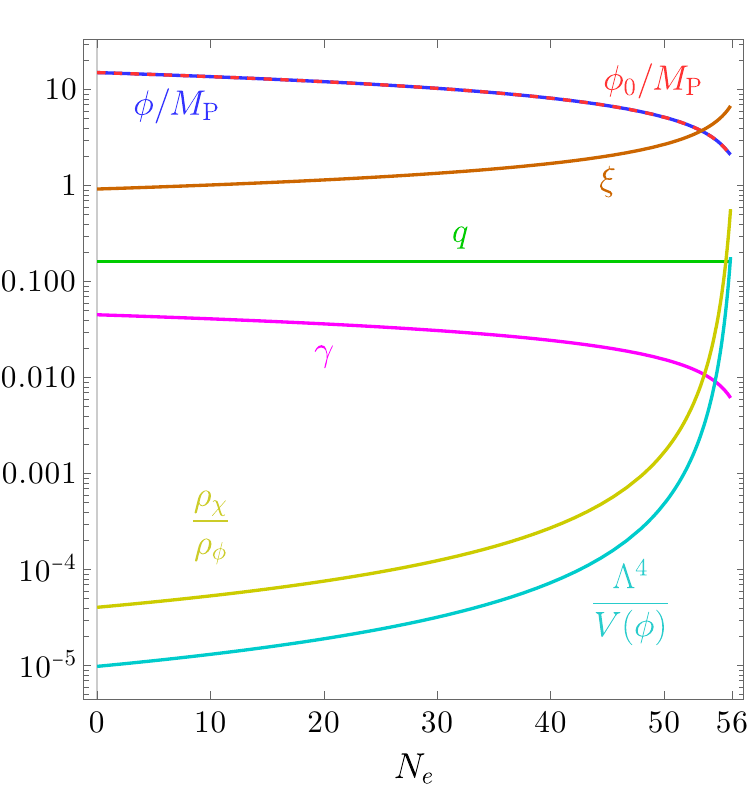}
			\caption{}
			\label{fig:field_params_quad_monomial}
		\end{subfigure}\hspace*{10pt}
		\begin{subfigure}{0.49\linewidth}
			\centering
			\includegraphics[width=0.977\linewidth]{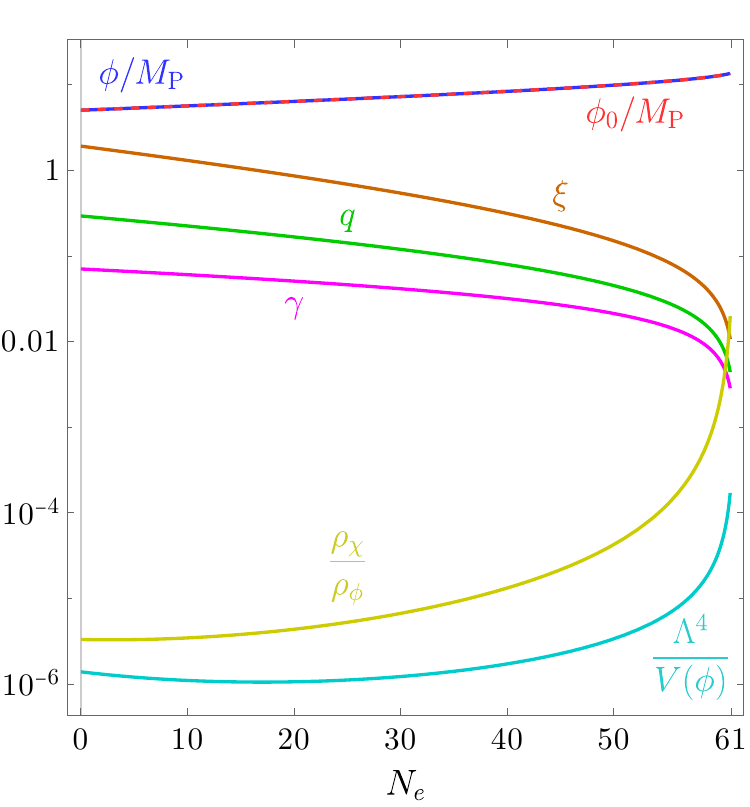}
			\caption{}
			\label{fig:field_params_quad_hilltop}
		\end{subfigure}
		\caption{Numerical evolution of relevant dynamical quantities during the final $\sim$60 $e$-folds of inflation for (a) the quadratic monomial potential with $\phi_i = 15\,M_\text{P}$ (taking $g = 0.35$ and $M= 3 \times 10^{15}$ GeV), and (b) the quadratic hilltop potential with $\phi_i = 5\,M_\text{P}$ and $\kappa = 10^{-2}$ (taking $g = 0.2$ and $M= 2\times 10^{15}$ GeV).}
		\label{fig:field_params}
	\end{figure}

In order to solve Eq.~\eqref{eq:EoM_phi1_phi_0}, we need to make a further approximation: following \cite{Flauger:2010ja}, we shall replace $\phi_0$ by a pivot value $\phi_*$ (taken to be the value of $\phi_0$ at the CMB pivot scale $k_*$) everywhere except in the argument of the sine on the right-hand side of the equation. This step basically assumes that the dynamics of $\phi_1$ is predominantly described by the oscillatory part of the equation, such that the coefficients of every term may be taken to be constant. In that case, the equation can be solved to find the full analytical solution for the field including backreaction
	\begin{equation} \label{eq:QCphi_sol}
		\phi(t) \:\: \simeq \:\: \phi_0(t) \:- \: \frac{3\, \Lambda^4 M}{4\, \epsilon_{V_*} V_*} \, \sin\left(\frac{2\, \phi_0(t)}{M}\right)~,
	\end{equation}
where the amplitude of the sine term is found to generally be quite small compared to $\phi_0$, which is typically $\mathcal{O}(M_\text{P})$ at CMB scales, while $\frac{\Lambda^4}{\epsilon_{V_*}\,V_*} \ll 1$. This furthermore shows that indeed $|\phi_1| \ll \frac{M}{2}$. In fact, plotting $\phi(t)$ against $\phi_0(t)$ for various families of potentials reveals that the two follow each other exceedingly closely, as illustrated in \Cref{fig:field_params} for the quadratic monomial and the quadratic hilltop potentials. Also plotted in this figure are the parameters $q$, $\gamma$ and $\xi$, as well as the ratios $\frac{\rho_\chi}{\rho_\phi}$ and $\frac{\Lambda^4}{V(\phi)}$.

		We note in particular the constancy of the $q$ parameter in the quadratic monomial case, as well as the growth of $\rho_\chi/\rho_\phi$ in the hilltop case despite $\xi$ strictly decreasing during slow-roll, being due to phase space effects as discussed in the previous section. In both cases, we find that $\rho_\chi$ approaches $\rho_\phi$ as inflation proceeds, the two densities becoming comparable in the monomial case at the end of the slow-roll period, which may potentially yield a smooth exit from inflation with no need for a separate reheating period.
		
	In spite of the reduced effect of the backreaction on the field dynamics, the slow-roll parameters from Eq.~\eqref{eq:QCslowrollparamshubble} for the same potentials exhibit considerable oscillations (see \Cref{fig:SR_params} below), which could naively prevent a slow-roll evolution.~This behaviour may be attributed to the very large oscillation frequency ($f \sim \frac{2H}{\gamma}$), such that only the average value of the slow-roll parameters impacts the field dynamics.~A similar effect has also been encountered in Ref. \cite{Ferraz:2023qia}.~Solving Eq.~\eqref{eq:BRfinalEoM} numerically and using this solution to determine the slow-roll parameters yields the same results. It appears that having an inflationary potential with a sufficiently small sinusoidal modulation does not have a great impact on the slow-roll dynamics (at the end of this subsection, we illustrate a situation where this no longer holds).

	To better understand this, we may follow a similar procedure to Ref. \cite{Ferraz:2023qia} and compute the average values of $\epsilon_\mathcal{H}$ and $\eta_\mathcal{H}$ within an oscillation. For this, we consider $	\mathcal{\dot H}( \phi) \simeq \frac{V^{-1/2}( \phi)}{2\sqrt{3} \,M_\text{P}} \big[ V_{,\phi_0}(\phi_0) - \frac{2\Lambda^4}{M}\sin \big(\frac{2 \phi_0}{M}\big) \big] \big(\dot \phi_0 + \dot \phi_1\big)$, where we have taken $\mathcal{V}^{-1/2}( \phi) \simeq V^{-1/2}( \phi)$ and $\mathcal{V}_{, \phi}( \phi) \simeq \mathcal{V}_{,\phi_0}( \phi_0)$, since $\Lambda^4 \ll V(\phi_0)$ and $ \phi \simeq \phi_0$. We also replace $\epsilon_*$ and $V_*$ in \eqref{eq:QCphi_sol} by $\epsilon_{V_0}$ and $V_0 = V(\phi_0)$, respectively, to a good approximation, which allows us to do some simplifications. After some algebra, we obtain
	\begin{equation} \label{eq:QCepsilon_analytical}
		\epsilon_\mathcal{H}  = \epsilon_{H_0} + \frac{ \text{sgn}(\dot \phi_0)\,\Lambda^4}{2\,V(\phi_0)}\left[\frac{4}{\gamma}\, \sin\left( \frac{2 \phi_0}{M} \right) + 3\, \cos\left( \frac{2 \phi_0}{M} \right) - \frac{3\, \Lambda^4}{\epsilon_{V_0} V(\phi_0)\, \gamma}\, \sin\left( \frac{4 \phi_0}{M} \right)\right]~.
	\end{equation}
Let us then compute the average value of the first term inside the square brackets over an oscillation period $T = \frac{2\pi}{2 |\dot \phi_0|/M} = \frac{\pi \gamma}{2H_0}$. Defining  $E_1(t) \equiv \frac{1}{6}\big(\frac{gM}{\pi M_\text{P}}\big)^2\,\frac{e\,^{\xi-q}}{\gamma^3}$, we find

	\begin{equation} \label{eq:avg_eps_1_computation}
		\begin{split}
			\left| \langle \Delta \epsilon_{H_0}^{(1)} \rangle_T \right|_\text{max}&\equiv  \left| \left\langle\frac{ 2\,\Lambda^4}{V(\phi_0)\,\gamma}\, \sin\left( \frac{2 \phi_0(t)}{M} \right) \right\rangle_T \right|\\
			&\simeq\:\: \left|\frac{1}{T} \int_{t_i-\frac{T}{2}}^{t_i+\frac{T}{2}} \left[E_1(t_i) + \dot E_1(t_i)\,(t-t_i)\, \sin \left(\frac{4H_0}{\gamma}\,t + \alpha \right)\right]\, dt\right|_\text{max} \\[5pt]
			&=\:\: \left|\frac{\dot E_1(t_i)}{T}\int_{t_i-\frac{T}{2}}^{t_i+\frac{T}{2}} t\, \sin \left(\frac{4H_0}{\gamma}\,t + \alpha \right)\, dt\right|_\text{max} \\
			&=\:\: \frac{1}{24} \left(\frac{gM}{\pi M_\text{P}}\right)^2  \frac{e\,^{\xi-q}}{\gamma^2} \,\Big|\xi \,(3\, \eta_{V_0}-2\,\epsilon_{V_0}) - 2q\,( \eta_{V_0}-\epsilon_{V_0}) - 3\,( \eta_{V_0}-2\,\epsilon_{V_0})\Big|  ~,
		\end{split}
	\end{equation}
where we used that $\phi_0(t) \simeq \phi_{0}(t_i) + \dot \phi_0(t_i)\, (t-t_i)$ in an oscillation period, given that $T \ll H_0^{-1}$, and the subscript ``max'' indicates that we computed the maximum value of this averaged quantity. A similar procedure can be followed for the two remaining terms, as well as for the $\eta_\mathcal{H}$ parameter. We plot in Figure \ref{fig:SR_params} the numerical evolution of the slow-roll parameters $\epsilon_\mathcal{H}$ and $\eta_\mathcal{H}$ for the two forms of the scalar potential considered previously, alongside the corresponding parameters without backreaction ($\epsilon_{H_0}$ and $\eta_{H_0}$) and including the averaged corrections from backreaction. As one can easily see, the averaged corrections from backreaction are significantly suppressed compared to the large oscillation amplitude of the numerical slow-roll parameters, thus explaining why particle production may have a mild impact on the field dynamics as obtained earlier when $\Lambda\ll V^{1/4}$.

	\begin{figure}[htb!]
	\centering
	\begin{subfigure}{0.49\linewidth}
		\centering
		\includegraphics[width=1\linewidth]{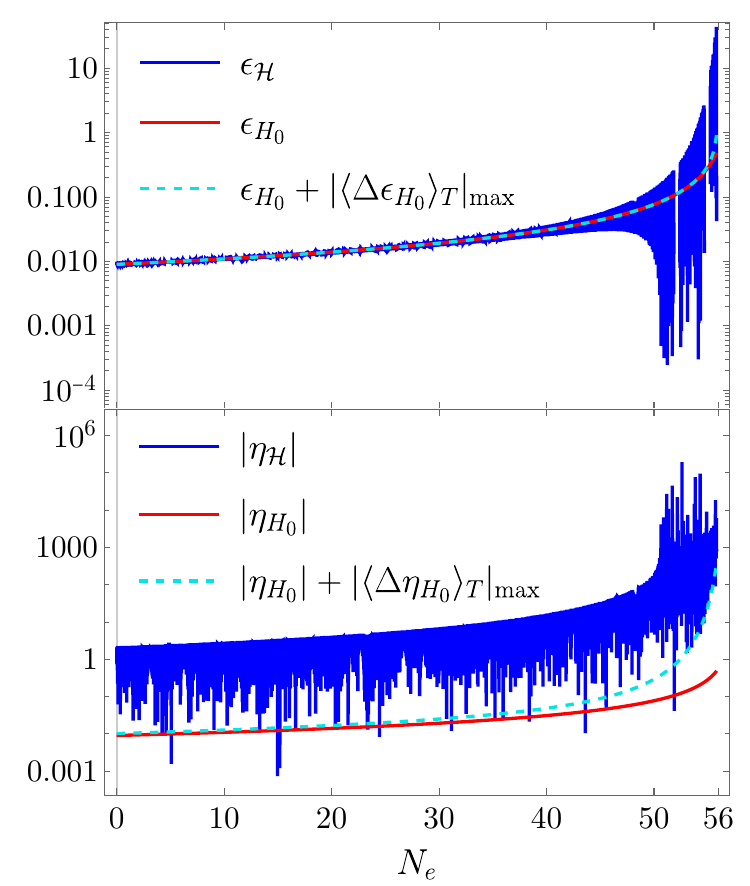}
		\caption{}
	\end{subfigure}\hspace*{10pt}
	\begin{subfigure}{0.49\linewidth}
		\centering
		\includegraphics[width=1\linewidth]{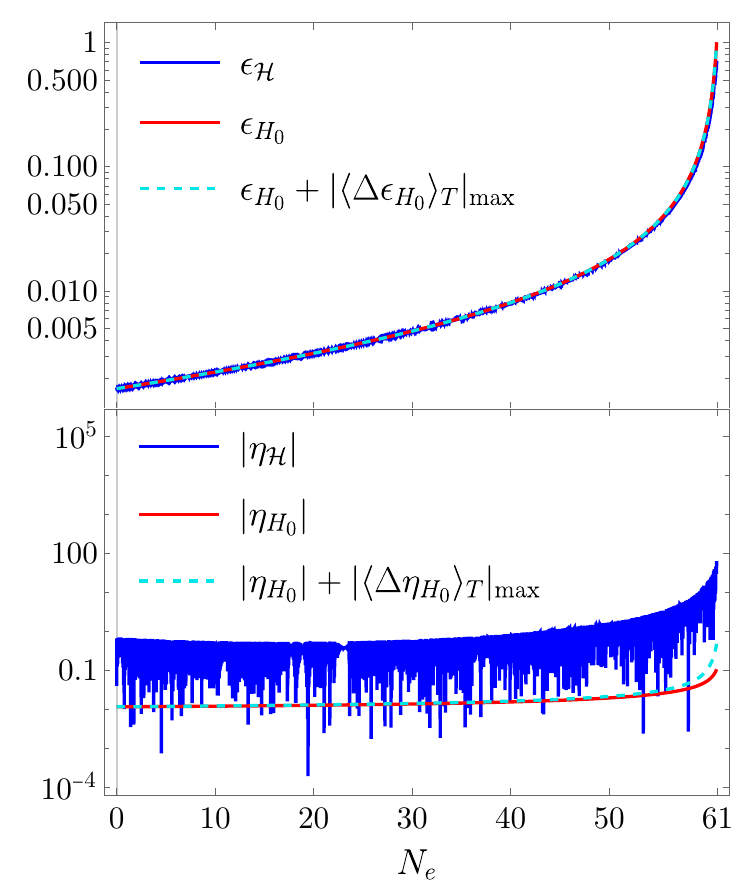}
		\caption{}
		
	\end{subfigure}
	\caption{Evolution of the Hubble slow-roll parameters including the effects of backreaction from particle production, before and after performing the averaging procedure discussed in the text, during the final $\sim$~60 $e$-folds of inflation for (a) the quadratic monomial potential with $\phi_i = 15\,M_\text{P}$ (taking $g = 0.35$ and $M= 3 \times 10^{15}$ GeV), and (b) the quadratic hilltop potential with $\phi_i = 5.8\,M_\text{P}$ and $\kappa =  10^{-2}$ (taking $g = 0.2$ and $M= 2\times 10^{15}$ GeV).}
	\label{fig:SR_params}
\end{figure}

The oscillations in the effective potential may also have an impact on the spectrum of curvature perturbations generated during inflation on super-horizon scales. Neglecting, for the time being, the potential impact on the dynamics of inflaton perturbations themselves, which we will study in the next sub-section, we may follow \cite{Flauger:2010ja}, which computed the corrections to the dimensionless curvature power spectrum for a scalar potential with oscillatory features, obtaining
	\begin{equation} \label{eq:QCpwr_spec_curv}
		 \Delta^2_\mathcal{R}(k) =  \Delta^2_{\mathcal{R}0}(k)\, \left[1 \: + \: \frac{3\, \Lambda^4 }{ \epsilon_{V_*} V_*} \sqrt{\frac{2 \pi}{\gamma_*}} \, \cos \left(\frac{2\, \phi_0(k)}{M}\right)\right]~,
	\end{equation}
where $ \Delta^2_{\mathcal{R}0}(k) = \Delta^2_{{\mathcal{R}0}}(k_*)\, \big(\frac{k}{k_*}\big)^{n_s-1}$, with $\Delta^2_{\mathcal{R}0}(k_*) = \big(\frac{H_*}{\dot \phi_*}\big)^2 \big(\frac{H_*}{2 \pi}\big)^2$, is the power spectrum obtained neglecting backreaction \cite{Dodelson:2020bqr,Riotto:2002yw}, $\phi_0(k)$ is the value of the uncorrected inflaton field when the mode with comoving momentum $k$ exits the horizon and $\gamma_* = \sqrt{\frac{2}{\epsilon_{V_*}}} \frac{M}{M_\text{P}}$. 

We thus see that oscillatory features in the potential lead to oscillatory corrections in the curvature power spectrum \cite{Planck:2018jri,Pahud:2008ae,Aich:2011qv,Flauger:2010ja}, in particular a fixed-amplitude sinusoidal oscillation\footnote{Recall that in deriving Eq.~\eqref{eq:QCpwr_spec_curv} we took $\Lambda^4\simeq $  constant. For a more general treatment taking into account the time-dependence of $\Lambda^4$, which we will not consider in this work, one could follow \cite{Flauger:2014ana}.}. It has been shown \cite{Pahud:2008ae, Aich:2011qv, Planck:2018jri} that compatibility with current CMB data requires $\delta n_s \equiv \frac{3\, \Lambda^4 }{ \epsilon_{V_*} V_*} \sqrt{\frac{2 \pi}{\gamma_*}}\lesssim 0.1$ (see in particular \cite{Planck:2018jri}). We have found that this condition is possible to attain in our scenario for all considered inflationary potentials, roughly corresponding to the alternative condition $\big| \frac{\Lambda^4}{V_*}\big| \lesssim 10^{-5}$ \cite{Aich:2011qv,Pahud:2008ae}, which in general means that the parametric resonance cannot too efficient at horizon crossing of CMB scales. We have, however, found parametric regimes (namely for hilltop potentials) for which $\big| \frac{\Lambda^4}{V(\phi)}\big| < 10^{-5}$ and $\xi > 1$, as shown in \Cref{fig:field_params_quad_hilltop}.
	
Although we will focus on observationally viable scenarios with small backreaction, it should be noted, for completeness, that for $\Lambda\lesssim V^{1/4}$ (i.e.~strong backreaction) the oscillatory features in the potential may produce local minima in the potential where the inflaton field gets trapped, apparently leading to eternal inflation. However, one must note that these features are a consequence of resonant particle production, which naturally stops if the field's velocity vanishes.

	\subsection{Fluctuation dynamics}
	\label{subsec:quantum}
	
	Let us now focus on determining how particle production affects the evolution of inflaton fluctuations. Under the conditions described in the previous section, we may neglect the corrections to the background inflaton field's evolution, so that inflaton fluctuations follow the linearized equation of motion
	\begin{equation} \label{eq:BR_fluc_EoM}
		\Box\varphi +\left(V''(\phi_0)+\frac{1}{2}m_\chi^2(\phi_0)\langle\chi^2\rangle\right) \varphi  = 0~.
	\end{equation}
%
Let us consider rescaled Fourier mode $u_k(t) = a^{3/2}(t)\, \varphi_k(t)$. Focusing first on sub-horizon modes, we may use the slow-roll condition ($V''(\phi_0)\ll H^2$) to yield
	\begin{equation} \label{eq:BR_fluc_modefunction_mathieu}
		u_k'' \,+\, \left[A_k(z) -2\, q_\varphi \cos (2z)\right]\, u_k =0\,, \quad 	\begin{cases}
			A_k(z)  \simeq\left(\frac{k/a}{2H/\gamma}\right)^2 \\
			q_\varphi  = \frac{g^2}{8 \pi^2}\: e\,^{\xi-q}
		\end{cases}~,
	\end{equation}
in terms of the previously defined $z$ variable, which is again a Mathieu-like equation. We therefore conclude that $\chi$-particle production results in an oscillating mass for the inflaton modes, which may also be produced via a (narrow) parametric resonance.~We note, in particular, that $A_k(z)$ has the same form as in Eq.~\eqref{eq:chimodesMathieu}, whereas $q_\varphi$ depends exponentially on both $\xi$ and $q$. This means that inflaton particles are also relativistic while the corresponding modes are inside the resonance band.

We may thus employ analogous methods to analyze $\chi$- and $\varphi$-particle production, but it should be immediately clear that the latter, being a secondary effect, is less efficient. In fact, we find $q_\varphi\ll q$ in all allowed regions of parameter space, particularly due to the $\frac{g^2}{8\pi^2}$ suppression factor, recalling also that the resonance efficiency parameter cannot be too large, at least before the relevant CMB scales have crossed the horizon.  

Fo these reasons we will not perform a detailed computation of the number density of $\varphi$-particles, focusing instead on the effects on the power spectrum of curvature perturbations. Following a similar procedure to the one used in \Cref{sec:resonant_PP}, each mode is amplified by a factor $e^{\mu_k^\varphi z}$ after crossing the resonance band, while deep inside the horizon, with $\mu_k^\varphi z \equiv \int_{z_i}^{z_f} \mu_k^\varphi(z) \: dz$. Once the mode has exited the resonance band, the oscillatory term has, on average, a vanishing effect and therefore the solution is the same as in the absence of $\chi$-particle production. This means that the power spectrum on super-horizon scales is simply modified by a multiplicative factor $e^{\xi_\varphi}$, where $\xi_\varphi\simeq {\frac{\pi q_\varphi^2}{ 2\gamma}}$, corresponding to the total amplification of the mode after crossing the resonance band.

While this effect is generically quite mild, it may be important in precision calculations of observables such as the spectral index and tensor-to-scalar ratio.~In particular, the resulting shift in the scalar spectral index is given by
\begin{equation} \label{eq:QC_scalarspecindex_quantum}
			\delta n_s^{res} \simeq \xi_{ \varphi} \left[2\xi\, (3 \eta_V-2 \epsilon_V) - 4 q\, (\eta_V-\epsilon_V) - (\eta_V-2 \epsilon_V)\right]~,
	\end{equation}
while the tensor-to-scalar ratio is simply suppressed by $e^{-\xi_\varphi}$.~These modifications are, however, generically very small, as well as somewhat degenerate with the choice of scalar potential. For example, when CMB scales cross the horizon, $\xi_\varphi\sim 10^{-4}$ for the quadratic monomial and hilltop potentials in the cases illustrated in Figure \ref{fig:field_params}, although it may reach larger values towards the end of inflation (in particular for the monomial potentials). Nevertheless, the resonant amplification of inflaton fluctuations leads to deviations from the canonical inflationary consistency relation
\begin{equation} \label{consistency}
r = -8 n_t e^{-\xi_\varphi}~,
\end{equation}
where $n_t=-2\epsilon_V$ is the tensor spectral index, which could in principle be tested in the future if the effects of primordial B-mode polarization on the CMB fluctuation spectrum are identified and measured with sufficient precision. This assumes, of course, that the generation of primordial gravitational waves is not significantly affected by particle production, which we discuss in the next section.

	\section{Secondary gravitational waves}
	\label{sec:gravitational_waves}

The resonant production of $\chi$-particles may also source secondary gravitational waves, potentially yielding another signature of this novel particle production mechanism.~To compute the resulting spectrum, we start by considering the equation for tensor metric perturbations in the transverse-traceless (TT) gauge, with a source term corresponding to the energy-momentum tensor of the produced $\chi$-particles \cite{Cook:2011hg,Dodelson:2020bqr,Caprini:2018mtu, Guzzetti:2016mkm,Boyle:2005se}
\begin{equation} \label{eq:hTT_EoM}
	\ddot h_{ij}^\text{TT} + 3 H\, \dot h_{ij}^\text{TT} + \frac{k^2}{a^2}\, h_{ij}^\text{TT} \:=\: \frac{2}{M_\text{P}^2}\,a^{-2}\: [T_{ij}^\chi]^\text{TT}(t, \mathbf{k})~,
\end{equation}
where the Fourier-transformed stress-energy tensor $[T_{ij}^\chi]^\text{TT}(t, \mathbf{k})$ is given by
\begin{equation} \label{eq:SET_TT_projection}
	[T_{ij}^\chi]^\text{TT}(t, \mathbf{k})\:\, \equiv\:\, \Pi_{ij}^{lm}(\mathbf{k})\left(\partial_l \hat \chi\, \partial_m \hat \chi\right)_\mathbf{k} \:\,=\:\,  \Pi_{ij}^{lm}(\mathbf{k}) \int \frac{d^3 k'}{(2 \pi)^3}\:\: k'_l \,k'_m\, \hat \chi_\mathbf{k'}(t)\,\hat \chi_{\mathbf{k}-\mathbf{k'}}(t)~,
\end{equation}
with $\hat \chi_\mathbf{k}(t) \equiv \hat a_\mathbf{k}\,\chi_\mathbf{k}(t) + \hat a_\mathbf{-k}^\dagger\,\chi^*_\mathbf{-k}(t)$, where $\chi_\mathbf{k}(t)$ is a complex mode function. The TT projector is given by $\Pi_{ij}^{lm}(\mathbf{k}) \equiv P_i^l(\mathbf{k})\, P_j^m(\mathbf{k}) - \frac{1}{2}\, P_{ij}(\mathbf{k})\,P^{lm}(\mathbf{k})$, with $ P_{ij}(\mathbf{k}) \equiv \delta_{ij} - \frac{k_i k_j}{|\mathbf{k}|^2}$. 

Considering some initial time $t_0$ (which will not affect the final result), the particular solution of Eq.~\eqref{eq:hTT_EoM} can be written as
\begin{equation}\label{eq:particular_solution}
h_{ij}^\text{TT}(t,\mathbf{k}) = \frac{2}{M_\text{P}^2} \int_{t_0}^t dt' \:\, a^{-2}(t')\:\,[T_{ij}^\chi]^\text{TT}(t', \mathbf{k}) \:\, G(t, t'; k)~,
\end{equation}
where the Green's function $G(t,t';k)$ can be expressed in terms of linearly independent solutions of the homogeneous (i.e. sourceless) equation
\begin{eqnarray}\label{eq:homogeneous_solutions}
h^{(1)}(t, k)\!=\!\cos \left(\frac{k}{aH}\right) + \frac{k}{aH} \sin \left(\frac{k}{aH}\right)~,\quad
h^{(2)}(t, k)\!=\!\sin \left(\frac{k}{aH}\right) - \frac{k}{aH} \cos \left(\frac{k}{aH}\right)
\end{eqnarray}
such that
\begin{equation} \label{eq:Greens_function}
	G(t,t';k) \:=\: \frac{h^{(1)}(t',k)\,h^{(2)}(t,k)-h^{(1)}(t,k)\,h^{(2)}(t',k)}{W[h^{(1)}(t',k),h^{(2)}(t',k)]}	\:\theta(t-t')~,
\end{equation}
where $W[h^{(1)}(t',k),h^{(2)}(t',k)] = \dot h^{(1)}(t',k)\, h^{(2)}(t',k) - h^{(1)}(t',k) \,\dot h^{(2)}(t',k)$ is the Wronskian of the two solutions (dots denoting derivatives with respect to $t'$) \cite{Cook:2011hg}.

The contribution of the $\chi$-induced gravitational waves to the tensor power spectrum $\Delta_t^2$ can be computed as the following vacuum expectation value
\begin{eqnarray} \label{eq:PSTP_VEV}
\frac{2\pi^2}{k^3}\delta^3(\mathbf{k}-\mathbf{k'})\,\Delta^2_t(t,\mathbf{k})\!\! &\equiv &\!\langle h_{ij}^\text{TT}(t,\mathbf{k})\, [h^{ij}_\text{TT}]^\dagger(t,\mathbf{k'})\rangle_\text{c} \nonumber\\
		& = &\left(\frac{2}{M_\text{P}^2}\right)^2\, \Pi_{ij}^{ab}(\mathbf{k})\,\Pi_{cd}^{ij}(\mathbf{k'})  \int \frac{d^3 p}{(2 \pi)^3} \int \frac{d^3 p'}{(2 \pi)^3}\: p_a\,p_b\,p'^c\,p'^d  \!\!\\
		&\!\!\times&\!\!\!\!\int_{t_0}^t\! dt' \frac{G(t,t';k)}{a^2(t')}\! \int_{t_0}^t\!\! dt'' \frac{G(t,t'';k')}{a^2(t'')}
		 \langle \hat \chi_\mathbf{p}(t')\,\hat \chi_{\mathbf{k}-\mathbf{p}}(t')\,\hat \chi^\dagger_{\mathbf{k'}-\mathbf{p'}}(t'') \,\hat \chi^\dagger_\mathbf{p'}(t'') \rangle_\text{c}\nonumber~,
\end{eqnarray}
where the subscript ``c'' denotes the connected part of the correlator.

The vacuum expectation value is determined using the standard commutation relations $[\hat a_\mathbf{k}, \hat a^\dagger_\mathbf{k'}] = (2 \pi)^3\, \delta^3(\mathbf{k}-\mathbf{k'})$, and the definition of the vacuum state $\hat a_\mathbf{k} \ket{0} = 0$, yielding
\begin{eqnarray} \label{eq:PSTP_computation_VEV_calculated_TT_projected}
		\Delta_t^2(t,\mathbf{k})&=&\frac{2k^3}{\pi^2 M_\text{P}^4}\int d^3 p\: \left(|\mathbf{p}|^2 - \frac{(\mathbf{p}\cdot \mathbf{k})^2}{|\mathbf{k}|^2}\right)^2 
		\int_{t_0}^t dt' \: \frac{G(t,t';k)}{a^2(t')} \int_{t_0}^t dt'' \: \frac{G(t,t'';k)}{a^2(t'')} \nonumber\\
		&\times& \chi_\mathbf{p}(t') \, \chi^*_\mathbf{p}(t'')\, \chi_{\mathbf{k}-\mathbf{p}}(t')\, \chi^*_{\mathbf{k}-\mathbf{p}}(t'')	~.
\end{eqnarray}
Since the Green's function in Eq.~\eqref{eq:Greens_function} is suppressed for large values of $k$, we shall focus on the case $k \ll p$, corresponding to $p$ modes that became superhorizon after the $k$ modes. Thus, given that the $\chi$ mode functions only depend on the absolute value of the momentum, and performing the angular integration, we may write
\begin{equation} \label{eq:PSTP_computation_VEV_calculated_TT_projected_approx}
	\Delta_t^2(t,k)\simeq\frac{64\, k^3}{15\pi M_\text{P}^4}\, \int_0^{p_\text{max}} \!\!\!\!\!d p \,p^6 
 \left[\int_{t_0}^t dt' \: \frac{G(t,t';k)}{a^2(t')}\:\left[\chi_p(t')\right]^2\right]  \left[\int_{t_0}^t dt'' \: \frac{G(t,t'';k)}{a^2(t'')}\:\left[\chi_p^*(t'')\right]^2\right] ~,
\end{equation}
The value of $p_\text{max}$ is determined by the momentum of the last $\chi$-mode produced during inflation via the relation $p_\text{max} = \frac{2H}{\gamma}\,a(t_e)$, with $t_e$ denoting the end of inflation. Furthermore, expanding the mode functions in terms of Bogoliubov coefficients \cite{Kofman:1997yn,Kachelriess:2022jmh}, we obtain $\left[\chi_p(t)\right]^2 \simeq \frac{a^{-3}(t)}{\omega_p(t)}\,\alpha_p(t)\,\beta_p(t)$, where high-frequency terms have been neglected.

We may now recall that the comoving occupation number for a certain $p$ mode, $\tilde n_p(t)$, is essentially a step function $\tilde n_p(t) \:\simeq\: \tilde n_p(t_+(p) \rightarrow t_-(p))\, \theta(t - t_c(p)) \:=\: \frac{e^{\xi(p)} - 1}{2}\:\theta(t-t_c(p))$ \cite{Rosa:2007dr}, where $\tilde n_p(t_+(p) \rightarrow t_-(p)) = \frac{1}{2}\,(e^{\xi(p)} - 1)$, with $\xi(p) \equiv \xi(t_c(p)) \equiv \frac{\pi q^2(t_c(p))}{2 \gamma(t_c(p))}$, is the overall amplification of the occupation number as the $p$ mode crosses the resonance band, while $t_c(p)$ denotes the time at which the $p$ mode is at the centre of the resonance band. Recalling also that $|\beta_p(t)|^2 = \tilde n_p(t)$ and using the fact that $|\alpha_p(t)|^2-|\beta_p(t)|^2=1$, we conclude that
\begin{eqnarray}\label{eq:Bogoliubov_occ_number}
		\beta_p(t)\simeq \sqrt{\frac{e^{\xi(p)}-1}{2}}\theta(t-t_c(p)) e^{i \arg \beta_p}~, \quad
		\alpha_p(t) \simeq \sqrt{1+ \frac{e^{\xi(p)}-1}{2}\theta(t-t_c(p))} e^{i \arg \alpha_p} 
\end{eqnarray}
Assuming the phases to be time-independent, we find
\begin{equation} \label{eq:PSTP_computation_VEV_calculated_TT_projected_approx_step_function}
	\Delta_t^2(t,k) \:\simeq\:\frac{16\, k^3}{15\pi M_\text{P}^4}\, \int_0^{p_\text{max}} d p\: p^6 \:(e^{2 \xi(p)}-1)  \left[\int_{t_c(p)}^t dt' \: \frac{G(t,t';k)}{\omega_p(t')\,a^5(t')}\right]^2~,
\end{equation}
at which point we may take the time $t$ to be $t_e$. Moreover, it proves simpler to convert the cosmic time $(t)$ integration into a conformal time $(\tau)$ integration. The two temporal quantities are related via $d \tau = \frac{dt}{a(t)}$, from where $a(\tau) = - \frac{1}{H \tau}$, with $\tau_e$ marking the end of inflation. The Green's function from Eq.~\eqref{eq:Greens_function} then simply becomes
\begin{equation} \label{eq:Greens_function_conformal}
	\begin{split}
	G(\tau,\tau';k) =&\, -\frac{1}{H \tau'}\, \frac{k(\tau-\tau')\: \cos\left[k(\tau-\tau')\right] - (1+k^2 \tau \tau')\:\sin\left[k(\tau-\tau')\right] }{k^3 \tau'^2} \: \theta(\tau-\tau') \\
	\equiv &\:\, a(\tau')\:\,\tilde G(\tau,\tau';k)~,
	\end{split}
\end{equation}
while the mode frequency is written as $\omega_p(\tau) \:=\: \sqrt{p^2H^2\, \tau^2+g^2M^2} \:\equiv\: \frac{\tilde \omega_p(\tau)}{a(\tau)}$.

Rewriting Eq.~\eqref{eq:PSTP_computation_VEV_calculated_TT_projected_approx_step_function} in conformal time and substituting in all the appropriate expressions thus yields, defining $\Delta_t^2(k)\equiv \Delta_t^2(\tau=\tau_e,k)$,
\begin{eqnarray} \label{eq:PSTP_computation_VEV_calculated_TT_projected_approx_conformal}
		\Delta_t^2(k) \:&\simeq&\:  \frac{16\, k^3}{15\pi M_\text{P}^4}\, \int_0^{p_\text{max}} d p\: p^6\: (e^{2 \xi(p)}-1) \left[\int_{\tau_c(p)}^{\tau_e} \frac{d\tau'}{a^2(\tau')} \: \frac{\tilde G(\tau_e,\tau';k)}{\tilde \omega_p(\tau')}\right]^2 \nonumber \\[5pt]
		&\simeq&\: D(k)\, \int_0^{p_\text{max}} d p\: p^6 \, H^4(p)  \: (e^{2 \xi(p)}-1)  \\
		&\:\:\:\:\:\:\:\:&\times \left[\int_{\tau_c(p)}^{\tau_e} d\tau' \: \frac{k(\tau_e-\tau')\: \cos\left[k(\tau_e-\tau')\right] - (1+k^2 \tau_e \tau')\:\sin\left[k(\tau_e-\tau')\right] }{\sqrt{p^2 + \frac{g^2 M^2}{H^2(p)\, \tau'^2}}}\right]^2\nonumber ~,
\end{eqnarray}
where $\tau_c(p) = - \frac{2}{\gamma p}$ and $D(k) \equiv \frac{16}{15\pi M_\text{P}^4 k^3}$. Notice that we have factored out $H^2$ of the time integral, since this quantity varies very little during slow-roll; however, we still consider its adiabatic variation by evaluating it at $\tau_c(p)$, i.e. $H(p) \equiv H(\tau_c(p))$, so that each $p$ mode experiences a slightly different value of $H$ as it crosses the resonance band. Notice also that $\gamma$ has a similar dependence on $p$ as well, which we shall omit in the notation for simplicity; henceforth, $\gamma \equiv \gamma(\tau_c(p))$. Furthermore, it is clear that the integrand function of the conformal time integral is suppressed for $\tau' \sim \tau_e$, so that the integral is dominated by $|\tau'| \gg |\tau_e|$. We may thus approximate $\tilde \omega_p(\tau') \simeq p$ in Eq.~\eqref{eq:PSTP_computation_VEV_calculated_TT_projected_approx_conformal}, in which case the conformal time integral can be easily computed, leaving us only with a comoving momentum integral that requires the specification of an inflationary potential in order to be computed.

The integrand in \eqref{eq:PSTP_computation_VEV_calculated_TT_projected_approx_conformal} is sharply peaked around its maximum at the comoving momentum $p_0(k)\simeq 0.6\, \gamma^{-1}\,k \gg k$ (within the narrow resonance regime), and we may approximate the quantities that vary little during slow-roll by their values computed when this mode crosses the resonance band, for each value of $k$, e.g. $H\simeq H(p_0(k))\equiv H_k$. We then obtain at the end of inflation
\begin{equation} \label{eq:PSTP_computation_VEV_calculated_TT_projected_approx_time_integration_p0}
	\begin{split}
		\Delta_t^2(k) \:\simeq\:&\,\frac{4 \,D(k)}{k^2}\:H_k^4 \: (e^{2 \xi_k}-1) \, \int_0^{-\frac{2}{\gamma_k\, \tau_e}} d p\, p^4 \\
		& \times \Bigg[ -1-\frac{k^2 \tau_e^2}{2} + \left(1- \frac{k}{\gamma_k\, p}\,k \tau_e \right)\,\cos \left(k \tau_e+\frac{2 k }{\gamma_k\, p}\right) \\
		&\quad\:\:\:\,+  \left(k\tau_e+\frac{k}{\gamma_k\, p}\right)\, \sin \left(k \tau_e+\frac{2k}{\gamma_k\, p}\right) \Bigg]^2~.
	\end{split}
\end{equation}
Since we are interested in modes that became super-horizon during the last $\sim60$ $e$-folds of inflation, $k\tau_e\ll 1$, we may perform the momentum integration to yield
\begin{equation} \label{eq:PSTP_computation_VEV_calculated_TT_projected_approx_final}
	\Delta_t^2(k) \:\simeq\: \frac{128}{225}\, \frac{e^{2 \xi_k} -1}{\gamma_k^5} \frac{H_k^4}{M_\text{P}^4} ~,
\end{equation}
which must be added to the primary (vacuum) contribution from inflation \cite{Dodelson:2020bqr,Riotto:2002yw}. This clearly shows that the GWs induced by $\chi$ production preserve the near scale invariance of the power spectrum, as would be expected from the nature of the resonance: since there is always a mode inside the resonance band at any given time during inflation and the $\chi$ quanta are continuously being produced, no particular comoving momentum scale is singled out in the GW spectrum. Specifically, $p_0$ is a $k$-dependent quantity, corresponding to the $\chi$ mode that contributes the most to the production of GWs of frequency $k$. As such, there does exist some degree of $k$ dependence, as the Hubble parameter, as well as the resonance parameters, all change adiabatically throughout inflation. In fact, this provides an additional observable to test our model.

\begin{figure}[htb!]
	\centering
	\begin{subfigure}{0.48\linewidth}
		\centering
		\includegraphics[width=1\linewidth]{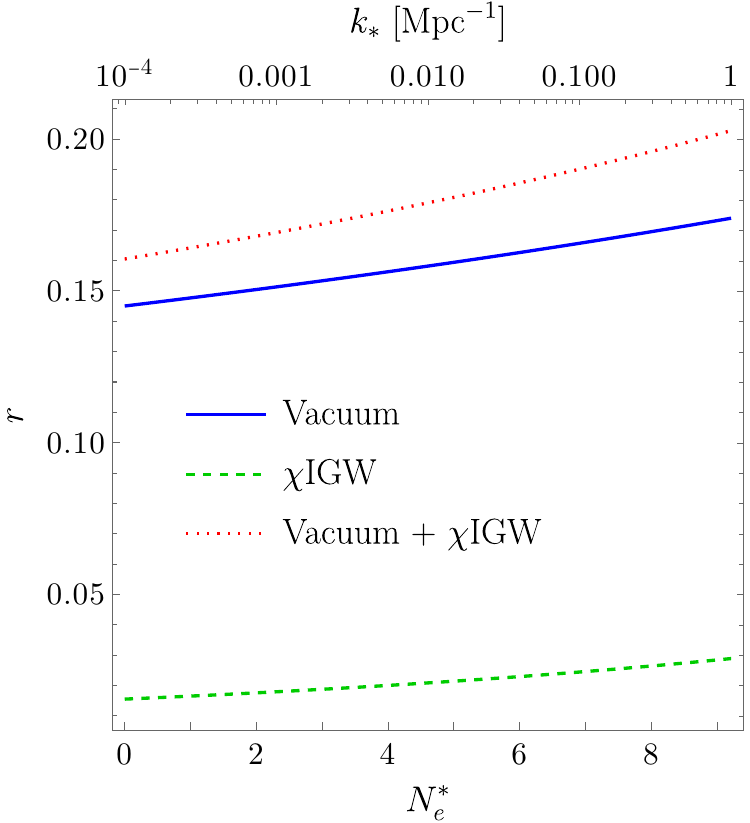}
		\caption{}
		\label{fig:powerspectens_quad_monomial}
	\end{subfigure}\hspace*{10pt}
	\begin{subfigure}{0.48\linewidth}
		\centering
		\includegraphics[width=1\linewidth]{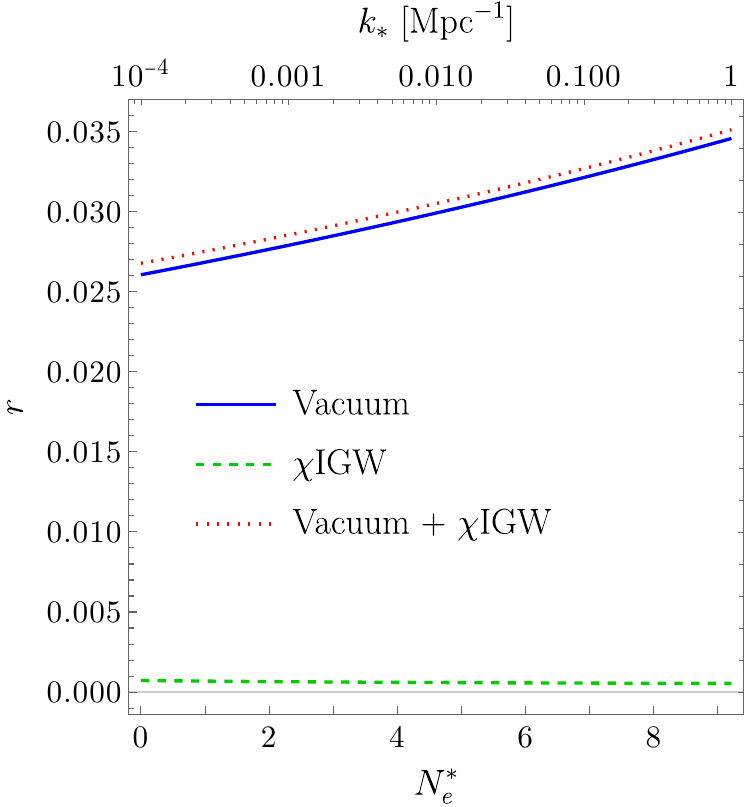}
		\caption{}
		\label{fig:powerspectens_quad_hilltop}
	\end{subfigure}
	\caption{Contribution of $\chi$-induced GWs ($\chi$IGW) to the tensor-to-scalar ratio at CMB scales (setting $N_e^*=0$ at $k_* = 10^{-4}$ Mpc$^{-1}$ \cite{Planck:2018jri}) for (a) the quadratic monomial potential with $\phi_i = 15\,M_\text{P}$ (taking $g = 0.35$ and $M= 3 \times 10^{15}$ GeV), and (b) the quadratic hilltop potential with $\phi_i = 5\,M_\text{P}$ and $\kappa =  10^{-2}$ (taking $g = 0.2$ and $M= 2\times 10^{15}$ GeV).}
	\label{fig:powerspectens}
\end{figure}

As can be seen in \Cref{fig:powerspectens}, the contribution of $\chi$IGWs to the tensor-to-scalar ratio is more significant for the quadratic monomial potential then for the quadratic hilltop potential, noting that only the latter is in agreement with observational data \cite{Planck:2018jri,ACT:2025tim}. 
For the hilltop case under consideration, we obtain a $\simeq 30\%$ change in the tensor spectral tilt $n_t$, which is the main modification to the consistency relation in \eqref{consistency}, yielding deviations of $\simeq 20\%$ from unity in $-r/8 n_t$, and which could potentially be probed with future CMB observations reaching this level of precision. However, the exponential dependence of Eq.~\eqref{eq:PSTP_computation_VEV_calculated_TT_projected_approx_final} on the evolution of the resonance efficiency parameter $\xi_k$ makes this result rather dependent on the form of the inflationary potential.

	\section{Discussion and Conclusion}
	\label{sec:discussion_conclusion}

In this work, we have shown that resonant particle production can occur during inflation and not only in the subsequent preheating period. In both cases a (narrow) parametric resonance occurs due to the oscillation of the mass of the produced particles, but in our proposed scenario this is due to the particle mass being itself an oscillatory function of the slowly rolling inflaton field. Interestingly, the underlying quantum field theory scenario is akin to the Warm Little Inflaton scenario that successfully realizes a nearly-thermal particle production during inflation, but we have demonstrated that it may also result in non-thermal particle production.

Much like in warm inflation, in our scenario a slowly varying particle density is sustained by resonant particle production, essentially since there are always field modes crossing the (first) resonance band and being exponentially amplified, which compensates for the effect of the quasi-exponential expansion, which dilutes the previously produced particles. The particle spectrum is, however, at all times sharply peaked around the physical momentum $p_c=\frac{2H}{\gamma} \gg H$, corresponding to the mode at the center of the resonance band. The particles are relativistic when produced, so we may refer to a non-thermal radiation bath being sustained in this preheated inflation scenario.

We have demonstrated that, at least in the narrow resonance regime, the radiation density may remain sub-dominant compared to the inflaton's potential energy for a wide range of parameters, typically corresponding to particle masses close the GUT scale. We have also examined the impact of particle production on the evolution of the background inflaton field, finding that the leading (Hartree) backreaction term in the effective potential has an oscillatory nature, which may produce large oscillations in the slow-roll parameters. These are, however, significantly suppressed since only their average effect impacts the field's evolution so that backreaction may be neglected throughout 50-60 $e$-folds of inflation in a broad range of mass scales and coupling constant values.

The oscillations in the effective potential naturally produce oscillations in the spectrum of scalar curvature perturbations on super-horizon scales, and we have discussed the parametric constraints imposed by CMB data, which are similar to e.g.~axion monodromy models. These constraints generically imply that the parametric resonance cannot be too efficient, at least when CMB scales cross the horizon during inflation. Another consequence of these oscillations is a secondary resonance leading to the production of inflaton quanta, and therefore amplifying the curvature power spectrum. However, since $\chi$-particle production cannot be too efficient, this secondary resonance has a negligible effect, at least for the forms of the scalar potential that we have considered as working examples. These two effects constitute, nevertheless, two testable predictions of the preheated inflation scenario.

We note, however, that the oscillations in the effective potential may be fully removed by considering the resonant production of two particle species, with masses oscillating with a $\frac{\pi}{2}$ phase shift (e.g.~a sine and a cosine). This can be achieved by imposing a discrete interchange symmetry on the underlying complex scalar fields, in addition to the broken U(1) gauge symmetry. Since this corresponds to a $\mathbb{Z}_2$ symmetry for the inflaton field, this would lead to a stable inflaton remnant at late times that could e.g.~account for dark matter \cite{Rosa:2018iff} or dark energy \cite{Rosa:2019jci} as shown in the context of the Warm Little Inflaton setup.

A final interesting and potentially testable consequence of resonant particle production during inflation that we have investigated is the production of secondary gravitational waves, sourced by the produced $\chi$-scalar particles. We have shown, in particular, that this may lead to small but nevertheless potentially meaurable changes in the tensor power spectrum that can be probed with future CMB experiments, as well as possibly directly with future gravitational wave detectors.

Our analysis examined only the most important features of this new form of particle production during inflation, and we aim to extend it in the future to investigate backreaction effects beyond the leading Hartree approximation, as well as possibly within the broad resonance regime, which may require dedicated lattice simulations akin to preheating, e.g.~using the CosmoLattice package \cite{Figueroa:2021yhd} or other similar tools. It would also be interesting to include interactions between the produced particles and the Standard Model degrees of freedom, alongside possible self-interactions, to study their potential thermalization either during or at the end of inflation, and hence examine in detail the ``graceful exit'' from inflation into the standard radiation era. We envisage, in particular, scenarios where the resonant production of right-handed (s)neutrinos along the lines proposed in \cite{Levy:2020zfo}, and which may also require studying fermion production, which is quite different from the bosonic case due to Pauli blocking \cite{Greene:2000ew}. We thus hope that this work motivates further studies of resonant particle production during inflation and of its observational impact.

	\acknowledgments
This work was supported by national funds by FCT - Funda\c{c}\~ao para a Ci\^encia e Tecnologia, I.P., through the research projects with DOI 10.54499/UIDB/04564/2020, \\10.54499/UIDP/04564/2020, 10.54499/CERN/FIS-PAR/0027/2021, and by the project 10.54499/2024.00252.CERN funded by measure RE-C06-i06.m02 – ``Reinforcement 
of funding for International Partnerships in Science, Technology and Innovation'' of 
the Recovery and Resilience Plan - RRP, within the framework of the financing contract signed between the Recover Portugal Mission Structure (EMRP) and the Foundation for Science and Technology I.P.
(FCT), as an intermediate beneficiary. D.S.G. was also partially supported by the Calouste Gulbenkian Foundation, under the Gulbenkian New Talents Scholarships.

%


\end{document}